\documentclass[prc,amssymb,amsmath,onecolumn,preprint]{revtex4-1}

\usepackage{graphicx}
\usepackage{caption}
\usepackage{subcaption}
\usepackage{amsmath}
\usepackage{epstopdf}
\usepackage{amsfonts,amssymb}
\usepackage[utf8]{inputenc}
\usepackage{pstricks}
\usepackage{color}
\usepackage{hyperref}
\usepackage[toc]{appendix}

\usepackage{color, colortbl}
\definecolor{Gray}{gray}{0.9}
\usepackage{float}
\usepackage{tikz-feynman}
\usepackage{forest}
\usetikzlibrary{shapes,arrows,positioning,automata,backgrounds,calc,er,patterns}
\usepackage{tikz-feynman}

\usepackage{xcolor}
\usepackage{amssymb}
\usepackage{float}
\usepackage{txfonts}
\usepackage{feynmp-auto}
\usepackage{braket}
\usepackage{tikz} 
\usepackage{simpler-wick}
\usetikzlibrary{shapes,arrows,positioning,automata,backgrounds,calc,er,patterns}
\usepackage[left=2.5cm,top=2.5cm,right=2.5cm,bottom=2.5cm]{geometry}
\tikzfeynmanset{compat=1.0.0}

\usepackage{verbatim}    
\usepackage{dcolumn}
\usepackage{bm}
\usepackage{epsfig}
\usepackage{braket}
\usepackage[normalem]{ulem} 
\usepackage{sidecap}

\newcommand{\be}{\begin{equation}}
\newcommand{\ee}{\end{equation}}
\newcommand{\ben}{\begin{eqnarray}}
\newcommand{\een}{\end{eqnarray}}

\def\keV{\mbox{ keV}} 
 
\def\GeV{\mbox{ GeV}}

\def\keV{\mbox{ keV}} 
 
\def\GeV{\mbox{ GeV}}

\newcommand{\pslash}{\not{\hbox{\kern-2.3pt $p$}}}
\newcommand{\pdslash}{\not{\hbox{\kern-2pt $\partial$}}}

\begin{document}
\title{Production of $ B \bar{B}$ bound state via $\Upsilon (4S)$ radiative decays}

\author{Andr\'e L. M. Britto}
\email{andrebritto@ufrb.edu.br}
\affiliation{ Centro de Ciências Exatas e Tecnológicas, Universidade Federal do Recôncavo da Bahia, R. Rui Barbosa, Cruz das Almas, 44380-000, Bahia, Brazil}

\author{Luciano M. Abreu}
\email{luciano.abreu@ufba.br}
\affiliation{ Instituto de F\'isica, Universidade Federal da Bahia,
Campus Universit\'ario de Ondina, 40170-115, Bahia, Brazil}

\begin{abstract}

Motivated by recent theoretical predictions about the existence of a $ B \bar{B}$ bound state (also denoted as $ X(10550) $), in this work we estimate the production of the $S$-wave $ B^+ B^-$ molecule via $\Upsilon (4S)$ radiative decays. In particular, we make use of effective Lagrangian approach and the compositeness condition to calculate the $ X(10550) $ production rate via $\Upsilon(4S)\rightarrow \gamma X(10550)$ decays employing triangle diagrams. 
Our results show that the partial decay width of this reaction is of the order of $0.5 - 192  \ \keV $ for a respective binding energy of $1 - 100 $ MeV, corresponding to a branching fraction of $ 10^{-5} - 10^{-3}$. These findings suggest that the existence of the $ X(10550) $ might be checked via the analysis of the mentioned decay in present and future experiments.

\end{abstract}
\maketitle

\section{INTRODUCTION}

In the last decades several new hadrons have been observed~\cite{Workman:2022ynf}, and many of them present unconventional properties that are incompatible with the quark model predictions~\cite{nora,yuan22,zhu,exo2}. Concerning their underlying structure, they have been interpreted by different configurations, e.g.: weakly bound hadron molecules, compact multiquark states, excited conventional hadrons, cusps engendered by kinematical singularities, glueballs, hybrids, etc., or even a superposition of some of them. The fact is that there is not an universal, consensual and compelling understanding on this point, and it remains as a topic of intense discussion. In order to establish criteria of distinction among these interpretations, observables like the masses, decay widths and production rates of these states have been studied both theoretically as well as experimentally. 
The most famous example is the $X(3872)$, the first observed exotic state in 2003~\cite{Workman:2022ynf,Belle:2003nnu}, with quantum numbers $I^G(J^{PC}) = 0^+(1^{++})$. Its intrinsic nature continues matter of dispute, and the most explored configurations are the weakly bound state of open charm mesons $(D\bar D^{*} +c.c.)$ and the $c \bar c q \bar q $ compact tetraquark~\cite{nora,yuan22,zhu,exo2}. 

With the observation of the $X(3872)$, a natural consequence in the scenario of meson molecule configuration was the investigation on the existence of its lightest partner, i.e. the $D\bar{D}$ state (also usually denoted as $X(3700)$ or $X(3720)$). It was predicted in the context of the coupled channel unitary approach~\cite{15}. In the sequence, this $ 0^+(0^{++})$ state was studied from distinct perspectives, namely: peak in the $D \bar{D}$ mass distribution of $e^+ e^- \to J/\psi D \bar{D}$ reactions~\cite{15bis,15bisbis}; via the pole structure of meson-meson interactions within the heavy meson effective theory~\cite{16,17,Ding:2020dio}; peak in the $\eta \eta^{\prime}$ mass distribution of the radiative decays of $\psi (3770), \psi (4040)$ and of the  $e^+ e^- \to J/\psi \eta \eta^{\prime}$~\cite{17bis}; peak in the $D^0 \bar{D}^0$ mass distribution of $\psi (3770) \rightarrow \gamma D^0 {\bar{D}}^0$ decay~\cite{Dai:2020yfu}; as a pole in the coupled $D \bar{D}, D_s \bar{D}_s$ scattering on lattice~\cite{18}; production in $\gamma \gamma \to D \bar{D}$ reactions~\cite{18bis}; its production in $B$ decays~\cite{Xie:2022lyw}; peak in the $\eta \eta$ mass distribution of $B^+ \to K^+ \eta \eta $ decay~\cite{Brandao:2023vyg}; its production in $\gamma \gamma \to D^+ D^-$ reactions seen in ultra-peripheral heavy ion collisions~\cite{Sobrinho:2024tre}; and so on. On experimental grounds, there are some searches reported in the literature. For example, Belle and BaBar Collaborations analyzed respectively the $e^+ e^- \to J/\psi D \bar{D}$ and  $e^+ e^- \to  D \bar{D}$ reactions~\cite{be1,be2,BaBar:2010jfn}, and although theoretical works claim that these data might be explained by the existence of the hidden charm scalar resonance~\cite{15bis,17bis,18bis,23}, there is not yet a consensus concerning its unequivocal observation.

Thus, by invoking heavy quark flavor symmetry one can naturally ask about the molecular partners in the bottom sector. Most importantly, however, the observation of the so-called $Z_b (10610)$ and $Z_b (10650)$ states with quantum numbers $ 1^+(1^+)$~\cite{Bondar,Adachi:2012im} indubitably enhanced the studies about hidden bottom meson molecules; see for instance the works~\cite{Bondar2, ClevenEPJ,Voloshin,Nieves2,Zhang,Sun,Yang,Ohkoda1,Ohkoda2,Li1,Li2,ClevenPRD,Ozpineci:2013zas,Ohkoda3,Dias,Kang,Huo,UFBa1,UFBaUSP1,Yang:2017prf,Ding:2020dio,Zhao:2021cvg}. 
In this sense, the community obviously has looked also at the case of the heavy quark symmetry partner of the  $X(3872)$, the $0^+(1^{++})$ state denoted as $X_b$ with a possible molecular configuration  $(B\bar B^{*} +c.c.)$. A lot of theoretical and experimental analyses have been developed exploring the similarities between these partner states (see e.g.~\cite{Ding:2020dio,Ozpineci:2013zas,Li:2014uia,Wu:2016dws,Yang:2017prf,Zhao:2021cvg,Wang:2023vkx,Aushev:2010bq,Belle:2014sys,Belle-II:2022xdi,CMS:2013ygz,ATLAS:2014mka,Liang:2019geg,Jia:2023pud,Liu:2024ets}), but no significant $X_b$ signals have been observed yet~\cite{Belle:2014sys,Belle-II:2022xdi,CMS:2013ygz,ATLAS:2014mka}. As noticed in Ref.~\cite{Wang:2023vkx}, at the current electron-positron colliders the direct observation of $X_b$ in hadronic decays is not likely because of its quantum numbers and large mass. Indeed, Belle (Belle-II) Collaboration has found no $X_b$ evidence in the search for $X_b \to \omega \Upsilon (1S)$~\cite{Belle:2014sys,Belle-II:2022xdi}. In addition, analyses of the CMS and ATLAS experiments at LHC based on samples of $ pp $ collisions at $\sqrt{s} = 8 $ TeV have searched for the $X_b$ decaying into $\Upsilon (1S) \pi^+ \pi^-$, and no significant excess above the background was observed~\cite{CMS:2013ygz,ATLAS:2014mka}, which might indicate that the isospin is conserved in this bottomonium system. Therefore, other possible channels have been proposed, and due to its high mass, a logical expectation is the $X_b$ production by means of the radiative decays of higher bottomonia. For example, in Refs.~\cite{Wu:2016dws,Wang:2023vkx} the $X_b$ production as a $(B\bar B^{*} +c.c.)$ molecule has been estimated small in the processes $\Upsilon (5S,6S) \to  \gamma X_b $, with a branching fraction of about $10^{-7}$. On the other hand, in Ref.~\cite{Liu:2024ets} the $X_b$ production via radiative transition of $\Upsilon (10753) $ has estimated to have a branching fraction by a factor about $10^{-3} - 10^{-2}$ higher that the former case, making it testable by future Belle-II experiments. 

Thus, benefiting from the discussion above, one can also focus on the bottomonium counterpart of the $X(3700)$ state, i.e. the $0^+(0^{++})$ state also denoted as $ X(10550) $, with a possible molecular configuration $B\bar{B}$. We mention that Ref.~\cite{AlFiky:2005jd} made use of an effective Lagrangian consistent with the heavy-quark and chiral symmetries, and argued that the existence
of a bound state in the $(D \bar D^{*} +c.c.)$ channel does not necessarily imply the  existence of a bound state in the $D\bar{D}$ or $B\bar{B}$ channels (see also an analysis of the $D D , \bar{B} \bar{B}$ cases in Ref.~\cite{Abreu:2022sra}). In contrast,  the meson-meson interaction has been analyzed in Ref.~\cite{Ozpineci:2013zas} via coupled channel unitary approach, combining the heavy quark spin symmetry and the dynamics of the local hidden gauge, and found the existence of a weakly $0^+(0^{++})$ $B\bar{B}$ bound state. In the sequence, other works have studied the possible existence and properties of the $ X(10550) $; the reader can consult for example Refs.~\cite{Yang:2017prf,Zhou:2018hlv,Liang:2019geg,Ding:2020dio}. 
Interestingly, we remark that, differently of the $X(3700)$, the analyses exploring the potential observation of the $ X(10550) $ via decays are very scarce. At the best of our knowledge, the studies available like~\cite{Liang:2019geg,Weng:2018ebv,Weng:2018ebv,Liu:2023gtx} investigate hadronic transitions with final states carrying $B^{(\ast)}\bar{B}^{(\ast)}$ or conventional bottomonia but without exploring the existence of the $ X(10550) $. 

Hence, considering the increasing interest on the exotic hadron spectroscopy and the search for more possible exotic states via estimation of relevant observables, as well as taking advantage of the similarities between the $0^+(0^{++})$ partner states in charmonium and bottomonium sectors discussed previously, in the present work we investigate the possible existence of the  $S$-wave  $ B^+ B^-$ bound state (here we continue denoting this charged component as $ X(10550) $), proposing a method to estimate its production via $\Upsilon (4S)$ radiative decays. In particular, we make use of effective Lagrangian approach and the compositeness condition to calculate the $ X(10550) $ production rate in $\Upsilon(4S)\rightarrow \gamma X(10550)$ decays employing triangle diagrams. 

This work is organized as follows. We introduce the formalism to calculate the amplitude associated to the triangle mechanism for the $\Upsilon(4S)\rightarrow \gamma X(10550) $ decay in Sec.~\ref{Framework}. Results and discussions are
given in Sec.~\ref{Results}, followed by concluding remarks in Sec.~\ref{Conclusions}.

\section{Formalism}
\label{Framework} 

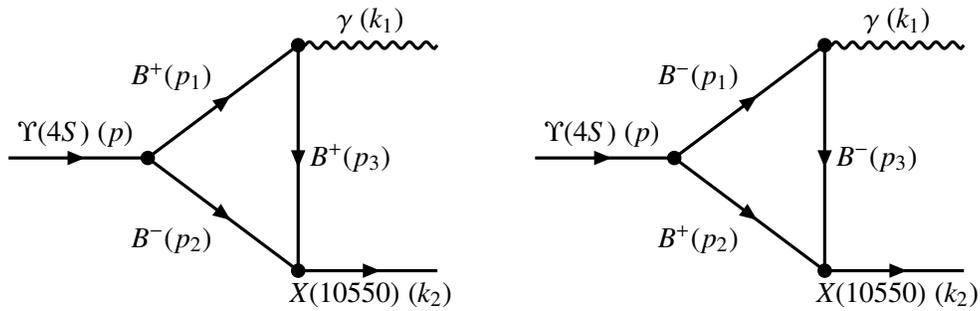
\begin{figure}[!htbp]
	\centering
\begin{tikzpicture}[very thick,q0/.style={->,semithick,yshift=5pt,shorten >=5pt,shorten <=5pt}]
\tikzfeynmanset{ every vertex = {dot} }
\begin{feynman}
    \vertex (a1){};
	\vertex[right=2.0cm of a1] (a2) ;
	\vertex[right=2.0cm of a2] (a3) {};
	\vertex[right=2.0cm of a3] (a4) {};
	\vertex[right=0.5cm of a4] (a41){};
	\vertex[right=0.5cm of a41] (a5){};
	\vertex[right=2.0cm of a5] (a6) ;
	\vertex[right=2.0cm of a6] (a7) {};
	\vertex[right=2.0cm of a7] (a8) {};
\vertex[below=1.5cm of a3] (c1);
\vertex[below=1.5cm of a4] (c2){};
\vertex[above=1.5cm of a3] (d1);
\vertex[above=1.5cm of a4] (d2){};
\vertex[below=1.5cm of a7] (c3);
\vertex[below=1.5cm of a8] (c4){};
\vertex[above=1.5cm of a7] (d3);
\vertex[above=1.5cm of a8] (d4){};
  \diagram* {
  (a1) --  [fermion, edge label=  {$\Upsilon(4S) \ (p)$}]  (a2), (a2) -- [fermion, edge label= {${B}^{+} (p_1)$}](d1), (d1) --  [photon, edge label= {$\gamma \ (k_1)$}](d2), (a2) --  [fermion, edge label'= {${B}^{-} (p_2)$}](c1), (c1) --[fermion, edge label'= {$X(10550) \ (k_2)$}]
   (c2), (d1) --  [fermion, edge label= {${B}^{+} (p_3)$}] (c1),  (a5) --  [fermion, edge label=  {$\Upsilon(4S) \ (p)$}] (a6), (a6) --[fermion, edge label= {${B}^{-} (p_1)$}] (d3), (d3) -- [photon, edge label= {$\gamma \ (k_1)$}](d4), (a6) --  [fermion, edge label'= {${B}^{+} (p_2)$}](c3), (c3) --  [fermion, edge label'= {$X(10550) \ (k_2)$}] (c4), (d3) -- [fermion, edge label= {${B}^{-} (p_3)$}](c3),
}; 
\end{feynman}
\end{tikzpicture}
\caption{Feynman diagrams for the production of the exotic state $ X(10550) $ via $\Upsilon (4S)$ radiative decays.}
\label{DIAG1}
\end{figure}

In what follows we describe the effective formalism used to evaluate the production of the so-called exotic state $ X(10550) $ via $\Upsilon (4S)$ radiative decays. Assuming this bound state as a $S$-wave $ B^+ B^-$  molecule (here denoted just as $B \bar{B}$) with quantum numbers $J^{PC} = 0^{++}$, then its production at the hadron level via the mentioned reactions can be described using the triangle diagrams depicted in Fig.~\ref{DIAG1}. To calculate the partial decay width of this reaction, we employ the effective Lagrangian approach.


We start by presenting the effective Lagrangian responsible for the interaction between the exotic state $ X(10550) $, here associated to the field $X$, and the $ B \bar{B} $ pair~\cite{Dong:2008gb}, 
\begin{align}
\mathcal{L}_{X B \bar{B}} = g_{X B \bar{B}} X(x) \int dy B(x+\omega_{\bar B B} y)\bar B(x-\omega_{ B \bar{B} } y)    \Phi(y), 
\label{LXBBbar}
\end{align}
where $y$ is the relative Jacobi coordinate and $\omega_{ij}= \frac{m_{i}}{m_i+m_{j}}$; since $m_{ B}  = m_{\bar{B} } $, we employ $\omega_{\bar B B} = 1/2 $ henceforth. $ \Phi(y)$ is the correlation function expressing the distribution of the two constituents hadrons in a molecule and also preventing the artificial growth of the amplitudes with energy. Its Fourier transform adopted here is the Gaussian function, 
\begin{align}
   \tilde{\Phi}(p^2) = e^{-p_E^2/\Lambda^2}
\label{Phi}
\end{align}
with $p_E$ being the Euclidean Jacobi momentum and $\Lambda$ a size parameter characterizing the distribution of the constituents inside the molecule. 

\begin{figure}[!htbp]
	\centering
\begin{tikzpicture}[very thick,q0/.style={->,semithick,yshift=5pt,shorten >=5pt,shorten <=5pt}]
\tikzfeynmanset{ every vertex = {dot} } 
\begin{feynman} 
\vertex (a1) {}; 
\vertex[right=2.5cm of a1] (a2); 
\vertex[right=2.5cm of a2] (a3); 
\vertex[right=2.5cm of a3] (a4) {}; 
\vertex[right=1.0cm of a2] (a5) {}; 
\vertex[above=1.0cm of a5] (a6) {}; 
\vertex[below=1.0cm of a6] (a7) {}; 
  \draw (3.75cm,0cm) circle (1.25cm);
\draw[->] (5cm,0cm) arc [radius = 1.25cm, start angle= 0, end angle= -270];
\draw[->] (5cm,0cm) arc [radius = 1.25cm, start angle= 0, end angle= -90];
\diagram* { (a1) -- [fermion, edge label'= {$X(10550) \ (k)$}] (a2),(a3) --[fermion, edge label'= {$X(10550) \ (k)$}] (a4) }; 
  \node[above] (a6) at (3.75cm,1.25cm) {$B(q_1)$};
  \node[below] (a7) at (3.75cm,-1.25cm) {$\bar B(q_2)$};
\end{feynman} 
\end{tikzpicture}

\caption{Feynman diagram contributing to the self-energy $\Sigma(k^2)$ of the $ X(10550) $ state.}
\label{DIAG2}
\end{figure}
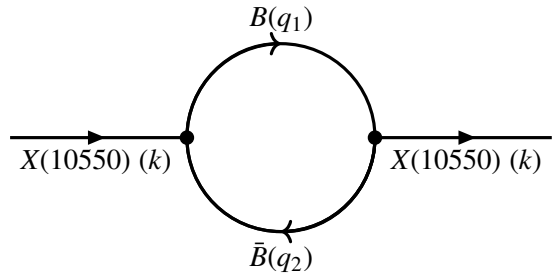

The coupling constant $g_{X B \bar{B}}$ can be estimated by the compositeness condition~\cite{Dong:2008gb,Weinberg:1962hj,Salam:1962ap,Hayashi:1967bjx,Ling:2021bir}. Accordingly, $g_{X B \bar{B}}$ is determined from the fact that the renormalization constant of the wave function associated to the composite state  $ X(10550) $ should be set equal to zero, i.e. 
\begin{equation}
Z_X = 1-\left.\frac{d\Sigma \left(k^2 \right)}{dk^2}\right|_{k^2=m_X^2}=0, 
\label{Zx}
\end{equation}
where $\Sigma(k^2)$ is the $ X(10550) $--self-energy, represented by the diagram of Fig.~\ref{DIAG2}. Here we define $m_X = m_B + m_{\bar{B}} - E_B $, with $E_B$ being the binding energy characterizing the state $ X(10550) $. 
The physical meaning of $Z_X = 0 $ is that the physical state is uniquely described by a bound state of its constituents, and as a consequence its mass and wave function must be renormalized due to the interaction of the $X$ with its constituents (see Refs.~\cite{Dong:2008gb,Weinberg:1962hj,Salam:1962ap,Hayashi:1967bjx,Ling:2021bir} for a more detailed discussion). 

The $X$--self-energy $\Sigma(k^2)$ represented in Fig.~(\ref{DIAG2}) can be expressed, after the use of some mathematical manipulations, Schwinger parametrization technique and Gaussian integration, in the form~\cite{Huang:2019qmw}
\begin{align}
   i\Sigma \left(k^2 \right)=i \frac{g^2}{16\pi^2}
    \int d\alpha d\beta
    \frac{1}{ z^2}
    \exp{\left\{
    -\frac{1}{\Lambda^2}
    \left[\frac{-k^2}{2}+m_B^2\alpha
    +(-k^2+ m_{\bar{B}}^2)\beta
    +\frac{\Delta^2}{4z}
    \right]
    \right\}}, 
\label{selfenergy1}
\end{align}
where $z = \alpha +\beta +2$ and $\Delta=-2k(1+\beta)$, with $\alpha , \beta$ being the Schwinger parameters. Thus, according to Eq.~(\ref{Zx}), the coupling constant $g_{X B \bar{B}}$ can be given by 
\begin{align}
    g_{X B \bar{B}}=
    \left\{
    \frac{1}{16\pi^2\Lambda^2}
     \int d\alpha d\beta
    \frac{1}{ z^2}\left(
    \beta+\frac{1}{2}-\frac{(1+\beta)^2}{\alpha+\beta+2}\right)
    \exp{\left[\frac{m_X^2}{\Lambda^2}\left(
    \beta+\frac{1}{2}-\frac{(1+\beta)^2}{\alpha+\beta+2}\right)\right]}
    \exp{\left[- \frac{m_B^2}{\Lambda^2}(\alpha+\beta)\right]}
    \right\}^{-\frac{1}{2}}.
\label{couplconst}
\end{align}

Beyond the effective Lagrangian given in Eq.~(\ref{LXBBbar}), to calculate the two-body decay via the triangle diagram shown in Fig.~\ref{DIAG1} one needs the effective Lagrangians responsible for the other vertices. The interactions between the  $\Upsilon (4S)$ and bottomed mesons are described by the following effective Lagrangian 
\begin{align}
    \mathcal{L}_{\Upsilon B \bar{B}} = - g_{\Upsilon B \bar{B}}\Upsilon_\mu\left(B \partial^\mu \bar{B} - 
    \bar{B} \partial^\mu B
    \right)
\label{YBBbar}
\end{align}
where $\Upsilon_\mu$ is the vector field associated to the $\Upsilon (4S)$. 
The coupling constant $g_{\Upsilon B \bar{B}}$ can be determined from the central values of the $\Upsilon (4S)$ total decay width, $\Gamma_ {\Upsilon(4S)} = 20.5 $ MeV, and the branching ratio $ \mathcal{B}_{\Upsilon(4S)\rightarrow B^+B^-} = \Gamma_{\Upsilon(4S)\rightarrow B^+B^-} / \Gamma_ {\Upsilon(4S)} = 51.4 \% $~\cite{Workman:2022ynf}. Therefore, we can relate the partial decay width $\Gamma _{\Upsilon(4S)\rightarrow B^+B^-}$ to  $g_{\Upsilon B \bar{B}}$  through the expression 
\begin{eqnarray}
 \Gamma_{\Upsilon(4S)\rightarrow B^+B^-} = \frac{1}{24\pi}\frac{|\vec{p}_B|^3}{ m^2_{\Upsilon}}  | g_{\Upsilon B \bar{B}} |^2 ,
  \label{Upsilondecaywidth}
\end{eqnarray}
where $|\vec{p}_B| = \lambda ^{1/2} \left(m^2_{\Upsilon},m^2_{B},m^2_{\bar{B}} \right) / (2 m_{\Upsilon})$ is the magnitude of the three-momentum of the $B$ meson in the rest frame of $\Upsilon$, with $ \lambda (a,b,c) = a^2 + b^2 + c^2 - 2 a b - 2 a c - 2 b c $ being the K\"allen function. With the use of experimental decay width and masses of the particles involved in this reaction reported in~\cite{Workman:2022ynf} we get  $g_{\Upsilon B \bar{B}} \sim 24$. 

The interaction between the pseudoscalar bottomed mesons and the photon is governed by the following Lagrangian, 
\begin{align}
    \mathcal{L}_{\gamma \bar{B} B}=-i e A_\mu
    \left( \bar{B} \partial^\mu B
    - 
      \partial^\mu \bar{B} \ B
    \right)
    +e^2 A^\mu A_\mu \bar{B} B , 
\label{LBBbarGamma}
\end{align}
coming from the usual scalar Quantum Electrodynamics. 

Thus, making use of the vertices discussed above, the amplitude of the radiative $\Upsilon (4S) \to  X(10550) \gamma $  decay given in Fig.~\ref{DIAG1} can be written as
\begin{align}
    \mathcal{M} &= - 2 i \  g_{X B \bar{B}} \ g_{\Upsilon B \bar{B}} \ e \  \epsilon^{(\gamma)}_\mu(k_1) \ \epsilon^{(\Upsilon)}_\nu (p)  \nonumber \\ 
    & \times
    \int \frac{d^4q}{(2\pi)^4} 
\frac{(2q-p )^\nu (2q-k_1  )^\mu}{\left[ (p-q)^2-m_B^2 + i \varepsilon \right] \left[ (q-k_1)^2- m_{\bar{B}}^2  + i \varepsilon \right] \left[ q^2- m_{\bar{B}}^2  + i \varepsilon \right]}
     \Phi\left(\frac{p+k_1}{2}-q\right) , 
    \label{ampl1}
´\end{align}
where the factor 2 in the right hand side of the first line comes from the fact that the two diagrams in Fig.~\ref{DIAG1} give equal contributions. Then, following~\cite{Aceti:2012cb,Dai:2020yfu}, $\mathcal{M} $ in Eq.~(\ref{ampl1}) can be expressed concisely as
\begin{align}
    \mathcal{M} &= - i \ \epsilon^{(\gamma)}_\mu(k_1) \ \epsilon^{(\Upsilon)}_\nu (p) 
     \left[ a \ g^{\mu\nu}+b \ k_1^\mu k_1^\nu+c \ k_1^\mu p^\nu +d\  k_1^\nu p^\mu + e \ p^\mu p^\nu \right] . 
    \label{ampl2}
\end{align}
By employing the transversality conditions $\epsilon^{(\gamma)}_\mu(k_1)  k_1^{\mu} = 0$ and  $\epsilon^{(\Upsilon)}_\nu (p)  p^{\nu} = 0$, only the terms carrying the coefficients $a$ and $d$ need to be calculated. Besides, invoking the Ward identity, which is equivalent to replace $ \epsilon^{(\gamma)}_\mu(k_1)$ by $k_{1\mu}$ in Eq.~(\ref{ampl2}) and requiring $k_{1\mu} \left[ a \ g^{\mu\nu} +d \ k_1^\nu p^\mu \right] = 0 $~\cite{Dai:2020yfu}, we have the relationship $a = - d \ k_1 \cdot p$. In addition, considering the Coulomb gauge, i.e. $ \epsilon^{(\gamma)}_0 = 0 $ and $ \epsilon^{(\gamma)}_i(k_1) k_1^i = 0 $, in the $\Upsilon(4S)$ rest frame the term $ \epsilon^{(\gamma)}_i(k_1) \ \epsilon^{(\Upsilon)}_j (p) \ d \ k_1^j \ p^i  $ vanishes. As a consequence, the amplitude takes the simplified form  
\begin{align} 
     \mathcal{M} &= i \ \epsilon^{(\gamma)}_\mu(k_1) \ \epsilon^{(\Upsilon) \mu} (p) \ d \ (k_1 \cdot p) .
    \label{ampl3}
\end{align}
The $d$ coefficient is obtained from Eq.~(\ref{ampl1}) and can be written by using the Schwinger parametrization as~\cite{Dai:2020yfu}
\begin{align}
    & d = -i \ g_{X B \bar{B}} \ g_{\Upsilon B \bar{B}} \ e \
    \frac{1}{16 \pi^2\Lambda^2 \ }  \nonumber \\ 
    & \times
    \int   d\alpha  d\beta  d\gamma
    \frac{1}{\tilde{z}^2}
\exp{\left\{\frac{1}{\Lambda^2}
\left[
   \left(-(\alpha  +\beta  +\gamma) \left(m_B^2-\frac{k_2^2}{4}\right)+k_1k_2\gamma 
   \right)-\frac{\tilde{\Delta}^2}{4 \tilde{z}}
\right]  
\right\}}, 
\label{dcoeff}
\end{align}
with $\tilde{z}=(\alpha +\beta +\gamma +1)$ and $\tilde{\Delta}= k_2(-\alpha+\beta+\gamma)+2k_1\gamma$. 

Finally, with all these ingredients discussed above, the partial decay width for the $\Upsilon (4S)$ radiative decay producing the exotic state $ X(10550) $ reads 
\begin{eqnarray}
 \Gamma_{\Upsilon(4S)\rightarrow \gamma X(10550) } = \frac{1}{8\pi}\frac{|\vec{k}_1|}{ m^2_{\Upsilon}} \sum \Bar{\sum} |  \mathcal{M} |^2,
  \label{partialdecaywidth}
\end{eqnarray}
where $|\vec{k}_1|$ is the magnitude of the three-momentum of the photon in the rest frame of $\Upsilon(4S)$, and $ \sum \Bar{\sum}$ represents the sum over the polarizations of the final state and average over the polarizations of the $\Upsilon (4S)$.

\section{Results}
\label{Results}

\begin{figure}[!htbp]
	\centering
\includegraphics[{width=8.0cm}]{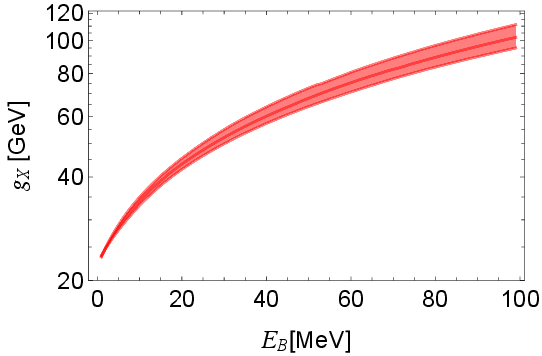} \\
\caption{Plot of the coupling constant $g_{X B \bar{B}}$ related to the molecular $X(10550)$ state and its  meson compoments $B , \bar{B}$ as a function of the binding energy $E_B$.  The band denotes the uncertainties coming from the values of the size parameter in the range $0.9\Lambda - 1.1 \Lambda$. }
\label{fig-couplconst}
\end{figure}

First, we present the estimation for the coupling constant $g_{X B \bar{B}}$ related to the $X(10550)$ state and its  meson compoments $B , \bar{B}$. As in other works (e.g. Refs.~\cite{Huang:2019qmw,Ling:2021lmq}), we use $\Lambda = 1 $ GeV. However, to take into account the uncertainties inherent in the approach we show the results within the range $0.9\Lambda - 1.1 \Lambda$. From  Eq.~(\ref{couplconst})  one can see that $g_{X B \bar{B}}$ is dependent on the mass of the bound state, and therefore on its binding energy. Noticing that in the literature (see for example Refs.~\cite{Ozpineci:2013zas,Ding:2020dio}) this state is predicted with a distinct $E_B$ in relation to the $B \bar{B}$ threshold, then in Fig.~\ref{fig-couplconst} we plot $g_{X B \bar{B}}$ obtained from the solution of Eq.~(\ref{couplconst}) as a function of the binding energy. We consider the range $ E_B \sim 1 - 100 $ MeV, corresponding to the bound state with mass $m_X \sim 10558 - 10458 $ MeV. It can be seen that  $g_{X B \bar{B}}$ acquires a bigger magnitude with the increasing $E_B$; in other words it growths as $m_X$ decreases. In Table~\ref{Tablecoupl-eb} we show explicitly the central values for the $g_{X B \bar{B}}$ for some specific binding energies. It should be mentioned that Ref.~\cite{Ozpineci:2013zas} has found a $B \bar{B}^{(I=0)}$ bound state using a coupled channel unitary approach, combining the heavy quark spin symmetry and the dynamics of the local hidden gauge for different cutoffs (and therefore different pole locations) in the same range of binding energy that we consider here, and reported values of the coupling in the same order of those for $g_{X B \bar{B}}$ shown in Table~\ref{Tablecoupl-eb}, although a direct comparison is not possible because Ref.~\cite{Ozpineci:2013zas} worked with potentials in the isospin basis, differently of our situation~\footnote{We assume that the values of the coupling reported in Tables V and VI of Ref.~\cite{Ozpineci:2013zas} are given in units of MeV.}.

\begin{figure}[!htbp]
	\centering
\includegraphics[{width=8.0cm}]{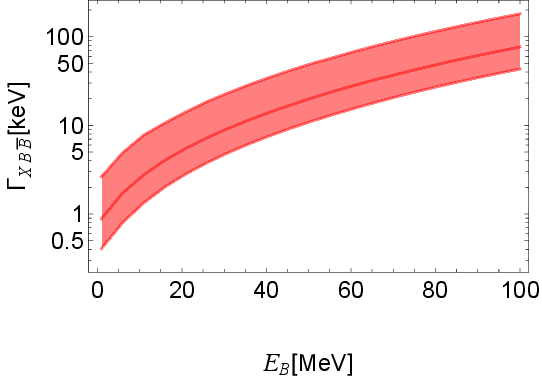} \\
\caption{Plot of the partial decay width $  \Gamma_{\Upsilon(4S)\rightarrow \gamma X(10550) } $ as a function of the binding energy $E_B$.  The band denotes the uncertainties coming from the values of the size parameter in the range $0.9\Lambda - 1.1 \Lambda$.}
\label{fig-decaywidth}
\end{figure}

Fig.~\ref{fig-decaywidth} exhibits the partial decay width for the the radiative decay $  \Upsilon(4S)\rightarrow \gamma X(10550) $, defined in Eq.~(\ref{partialdecaywidth}), as a function of the binding energy. It can be seen that with increasing $E_B$, the radiative decay width increases. This behavior comes essentially from the dependence of $g_{X B \bar{B}}$ with the binding energy previously discussed. In Table~\ref{Tablecoupl-eb} we show explicitly the central values for the decay width $  \Gamma_{\Upsilon(4S)\rightarrow \gamma X(10550) } $ and the branching ratio for some specific binding energies.  In particular, assuming the $ X(10550) $ molecule with a binding energy of $1 - 100 $ MeV, also within the range considered in Ref.~\cite{Ozpineci:2013zas}, i.e. $m_X \sim 10558 - 10458 $ MeV, the radiative decay width predicted is 
\begin{eqnarray}
 \Gamma_{\Upsilon(4S)\rightarrow \gamma X(10550) } \sim 0.5 - 192 \keV ,
  \label{partialdecaywidthEB}
\end{eqnarray}
which engenders a branching ratio of the order $ \mathcal{B}_{\Upsilon(4S)\rightarrow \gamma X(10550)} \sim 10^{-5} - 10^{-3} $. 
    This result indicates a relative large radiative width, suggesting a promising hunt for $X(10550)$ via the $\Upsilon(4S)\rightarrow \gamma X(10550)$ decay in updated Belle II experiments. This is the main finding of the present study. 

\begin{table}[!htbp]
\centering
\caption{Central values (considering $\Lambda = 1 \ \GeV$) of the coupling constant $g_{X B \bar{B}}$ related to the molecular $X(10550)$ state and its  meson components $B ,\bar{B}$, the decay width $  \Gamma_{\Upsilon(4S)\rightarrow \gamma X(10550) } $, and the branching ratio for some values of the binding energy $E_B$.}
\begin{tabular}{l|l |l |l}
\hline
\hline
$E_B$ [MeV]    & $g_{X B \bar{B}} $ [GeV] & $ \Gamma_{\Upsilon(4S)\rightarrow \gamma X(10550) }$  [keV] & $\mathcal{B}_{\Upsilon(4S)\rightarrow \gamma X(10550)} $  \\ \hline
5   &  28.50  &  1.55 & 7.55$\times 10^{-5}$\\
10  & 33.96   &   2.59   & 1.25$\times 10^{-4}$\\
25  & 47.94   &  6.90    &  3.37$\times 10^{-4}$   \\ 
50  &  67.87  &  19.87   &   9.70$\times 10^{-4}$ \\ 
75  &  85.85  &  42.34   &   2.06$\times 10^{-3}$ \\
100 & 102.73  &  76.79   &   3.75$\times 10^{-3}$ \\ \hline\hline
\end{tabular}
\label{Tablecoupl-eb}
\end{table}

A final remark is that the prediction of the radiative decay width predicted in Eq.~(  \ref{partialdecaywidthEB}) is (taking into account the uncertainties) of the same order of the one reported in Ref.~\cite{Liu:2024ets} for the production of the $X_b$, the heavy quark flavor symmetry counterpart of the $X(3782)$ in the bottomonium sector, via radiative transition of the $\Upsilon (10753)$, seen as a $S-D$ mixed state of the $\Upsilon(4S)$ and $\Upsilon_1 (3^3 D_1S)$, within a distinct framework based on a nonrelativistic effective field theory. Hence, these two findings corroborate the viewpoint that the radiative decays of $\Upsilon$ states might be an interesting ground for the study of new exotic states in the bottomonia sector.  

\section{Concluding remarks}
\label{Conclusions}

In this work we have proposed the search for the $ X(10550) $ state, assumed as a $S$-wave $(0^{++})$ $ B^+ B^-$ molecule, via $\Upsilon (4S)$ radiative decays. In this sense, the $ X(10550) $ production rate for the $\Upsilon(4S)\rightarrow \gamma X(10550)$ process, described by triangle diagrams, has been evaluated by making use of an Effective Lagrangian approach and the compositeness condition. 
The partial decay width of this reaction has been estimated to be of the order of $0.5 - 192 \ \keV $ for a respective range of binding energy of $1 - 100 $ MeV , corresponding to a branching fraction of $ 10^{-5} - 10^{-3}$. It is sufficiently large to check the existence of the $ X(10550) $ via the mentioned channel in present and future experiments. We hope that this finding may stimulate experimental initiatives in this direction. On that regard, an eventual observation of the $ X(10550) $ might represent another relevant piece in the puzzle of the exotic hadron spectroscopy and help us to shed more light on the intrinsic nature of the partner states related by the heavy quark symmetry.

\begin{acknowledgements}

We would like to thank Eulogio Oset for fruitful discussions. This work was partly supported  by the Brazilian agencies CNPq/FAPERJ under the Project INCT-Física Nuclear e Aplicações (Contract No. 464898/2014-5). The work of L.M.A. is partly supported by the Brazilian agency CNPq under Contracts  No. 400215/2022-5, 200567/2022-5 and 308299/2023-0.

\end{acknowledgements}


\end{document}